\documentclass[a4paper,12pt]{article}
\usepackage{epsfig,amsmath,amssymb,amsthm}

\usepackage{openbib}
\usepackage[round]{natbib}
\usepackage[ansinew]{inputenc}
\usepackage{graphicx}

\usepackage{enumerate}
\usepackage{lscape}
\usepackage{fullpage}
\usepackage{setspace}
\usepackage{enumitem}
\usepackage{subfigure}

\usepackage{rotating}

\relax

\newcommand{\mathe}{\mathsf{E}}

\usepackage{verbatim,color}

\def\boxit#1{\vbox{\hrule\hbox{\vrule\kern6pt
          \vbox{\kern6pt#1\kern6pt}\kern6pt\vrule}\hrule}}

\newcommand{\captionfonts}{\small}

\makeatletter
\long\def\@makecaption#1#2{%
  \vskip\abovecaptionskip
  \sbox\@tempboxa{{\captionfonts #1: #2}}%
  \ifdim \wd\@tempboxa >\hsize
    {\captionfonts #1: #2\par}
  \else
    \hbox to\hsize{\hfil\box\@tempboxa\hfil}%
  \fi
  \vskip\belowcaptionskip}
\makeatother
\title{\textsc{A nonparametric test for a constant correlation matrix}}

\author{\textsc{Dominik Wied}\footnote{TU Dortmund, Fakultät Statistik, 44221 Dortmund, Germany.\ Email: wied@statistik.tu-dortmund.de, Phone: +49 231 755 5419.\
Financial support by Deutsche Forschungsgemeinschaft (SFB 823, project A1) and useful comments by the editor Esfandiar Maasoumi, an associate editor, a referee, Matthias Arnold, Axel Bücher, Walter Krämer and Martin Wendler are gratefully acknowledged.} \\ \small \textit{TU Dortmund}}

\date{\today}

\begin{document}

\doublespacing

\parindent 0cm

\makeatletter
\def\@seccntformat#1{\csname the#1\endcsname. }
\def\section{\@startsection {section}{1}{\z@}{-3.5ex plus -1ex minus
    -.2ex}{1.3ex plus .2ex}{\center\large\sc}}
\def\subsection{\@startsection{subsection}{2}{\z@}{3.25ex plus 1ex minus .2ex}{-1em}{\normalsize\bf}}

\def\subsubsection{\@startsection{subsubsection}{3}{\z@}{3.25ex plus 1ex minus .2ex}{-1em}{\normalsize\it}}

\newtheorem{theorem}{Theorem}
\newtheorem{definition}{Definition}
\newtheorem{lemma}{Lemma}
\newtheorem{assumption}{Assumption}
\newtheorem{cor}{Corollary}
\theoremstyle{definition}
\newtheorem{example}{Example}
\newtheorem{remark}{Remark}

\maketitle

\begin{abstract}
We propose a nonparametric procedure to test for changes in correlation matrices at an unknown point in time.
The new test requires constant expectations and variances, but only mild assumptions on the serial dependence structure and has considerable power in finite samples.
We derive the asymptotic distribution under the null hypothesis of no change as well as local power results and apply the test to stock returns.
\end{abstract}



\newpage

\section{Introduction} \label{sec:introduction}

The Bravais-Pearson correlation coefficient is arguably the most widely used measure of dependence between random variables.
For financial time series, correlations among returns are for instance widely used in risk management.
However, there is compelling empirical evidence that
the correlation structure of financial returns cannot be assumed to be constant over time,
see e.g. \citet{krishan:2009}.
In particular, in periods of financial crisis, correlations often increase, a phenomenon which is sometimes referred to as ``Diversification Meltdown''.
As most often potential change points are not known a priori, practitioners are interested in testing correlation constancy in financial time series at an unknown point in time.

\citet{wied:2012} propose a nonparametric retrospective kernel-based correlation constancy test (referred to as KB-test in what follows)
and \citet{wiedgaleano:2012} propose a sequential monitoring procedure. These papers complement other approaches for related measures of dependence, e.g.\ for the whole covariance matrix (\citealp{aue:2009b}, Galeano and Pe\~na, 2007\nocite{galeano:2007}),
the copula (\citealp{na:2012}, \citealp{kramervk:2010}), Spearman's rho (\citealp{gaissler:2010}), Kendall's tau (\citealp{dehling:2012}), autocovariances in a linear process (\citealp{lee:2003}) and covariance operators in the context of functional data analysis (\citealp{fremdt:2012}).

In what follows, we stick to correlation.\ \citet{wied:2012} show that a correlation test can be more powerful than a covariance test when we have more than one change point in the covariance structure.

However, the KB-test only considers bivariate correlations, whereas in portfolio management, where we typically have more than two assets, constancy of the whole correlation matrix is of interest. In this context, it would be possible to perform several pairwise tests and to use a level correction like Bonferroni-Holm. However, in this paper, we consider the correlation matrix. In a simulation study, we see that the matrix-based test outperforms the Bonferroni-Holm approach in some (although not in all) situations. We extend the methodology from the KB-test to higher dimensions, but on the other hand keep its nonparametric and model-free approach. 

We consider the $\frac{p(p-1)}{2}$-vector of
successively calculated pairwise correlation coefficients and derive its limiting distribution with the functional delta method approach and some proof ideas from \citet{wied:2012}. We use a bootstrap approximation for a normalizing constant in order to approximate the asymptotic limit distribution of the test statistic.
This may be an alternative for the bivariate case as well.

In an application of this test to Value-at-Risk forecasts (\citealp{berens:2013}) it is seen that this proposed test might indeed be useful in practical situations. That is, it might be a promising approach to combine the well-known CCC (constant conditional correlation) and DCC (dynamic conditional correlation) model with this test for structural breaks in correlations.

The paper is organized as follows: In Section 2, we present the test statistic and derive the asymptotic null distribution, Section 3 deals with local power, Section 4 presents simulation evidence, Section 5 provides an empirical illustration and Section 6 a conclusion. All proofs are deferred to the appendix.

\section{The Fluctuation Test} \label{sec:retrotest}

Let ${\bf X}_t = \left( X_{1,t},X_{2,t},\ldots,X_{p,t}\right) $, $t\in \mathbb{Z}$, be a sequence of $p$-variate random vectors on a probability space $(\Omega,\mathfrak{A},\mathsf{P})$ with finite $4$-th moments and
(unconditional) correlation matrix $R_t = (\rho^{ij}_t)_{1 \leq i,j \leq p}$, where
\begin{equation*}
\rho^{ij}_{t}=\frac{\mathsf{Cov}(X_{i,t},X_{j,t})}{\sqrt{\mathsf{Var}(X_{i,t})\mathsf{%
Var}(X_{j,t})}}.
\end{equation*}
Furthermore, we call $||\cdot||_r$ the $L_r$-norm, $r > 0,$ and $D(I,\mathbb{R}^d), d \in \mathbb{N},$ the space of $d$-dimensional càdlàg functions on an interval $I \subseteq [0,1]$ (compare \citealp{billingsley:1968} and related literature for details).
We write $A \sim (m,n)$ for a matrix $A$ with $m$ rows and $n$ columns. Throughout the paper, we denote by $\rightarrow_d$ and $\rightarrow_p$ convergence in distribution and probability, respectively, of random variables or vectors. By $\Rightarrow_d$, we denote convergence in distribution of stochastic processes on a function space, which will be specified depending upon the situation, and with respect to the corresponding supremum norm.

For $T \in \mathbb{N}$, the hypothesis pair is given by $H_0: R_1 = \ldots = R_T$ vs. $H_1: \neg\ H_0$. Under $H_0$, we denote $\rho^{ij}_{t} =: \rho^{ij}$.

The ``preliminary version'' of the test statistic is given by

\begin{equation*}
Q_T := \max_{2 \leq k \leq T} \sum_{1 \leq i < j \leq p} \frac{k}{\sqrt{T}} \left| \hat \rho^{ij}_k - \hat \rho^{ij}_T \right| =:
\max_{2 \leq k \leq T} \frac{k}{\sqrt{T}} \left| \left| P_{k,T} \right| \right|_1,
\end{equation*}
where
\begin{align*}
\hat \rho^{ij}_k = \frac{\sum_{t=1}^k (X_{i,t} - \bar X_{i,k})(X_{j,t} - \bar X_{j,k})}{\sqrt{\sum_{t=1}^k (X_{i,t} - \bar X_{i,k})^2}\sqrt{\sum_{t=1}^k
(X_{j,t} - \bar X_{j,k})^2}}, \label{rhohat}
\end{align*}
$\bar X_{i,k} = \frac{1}{k} \sum_{t=1}^k X_{i,t},\ \bar X_{j,k} = \frac{1}{k} \sum_{t=1}^k X_{j,t}$ and $P_{k,T} = \left(\hat \rho^{ij}_k - \hat \rho^{ij}_T \right)_{1 \leq i < j \leq p}
\in \mathbb{R}^{\frac{p(p-1)}{2}}$.\footnote{Here and analogously in the following, the expression $1 \leq i < j \leq p$ for a vector means that the first entry or entries consist of the expressions for $i=1$, followed by the one(s) for $i=2$ and so on.}
The value $\hat \rho^{ij}_k$ is the empirical pairwise correlation coefficient for the random variables $X_i$ and $X_j$, calculated from the first $k$ observations. Thus, the test statistic compares the pairwise successively calculated correlation coefficients with the corresponding correlation coefficients calculated from the whole sample. The null hypothesis is rejected whenever $Q_T$ becomes too large, i.e., whenever at least one of these differences become too large over time or, equivalently, whenever the successively calculated correlation coefficients of at least one pair fluctuate too much over time. The weighting factor $\frac{k}{\sqrt{T}}$ serves for compensating the fact that the correlations are typically estimated better in the middle or in the end of sample compared to the beginning of the sample. We will see later on in the context of discussing the bootstrap approximation that it might be more convenient to use a slightly modified version of $Q_T$.

For deriving the limiting null distribution and local power results, some additional assumptions
are necessary. The following assumptions concern moments and serial dependencies of the random variables and correspond to (A1), (A2) and (A3)
in \citet{wied:2012}, adjusted for the multivariate case.

\begin{assumption}\label{limit}
For
\begin{equation*}
U_t := \begin{pmatrix} X^2_{1,t} & - & \mathsf{E}(X^2_{1,t}) \\
                                                        \vdots & & \vdots \\
                                                     X^2_{p,t} & - & \mathsf{E}(X^2_{p,t}) \\
                                                     X_{1,t} & - & \mathsf{E}(X_{1,t}) \\
                                                        \vdots & & \vdots \\
                                                     X_{p,t} & - & \mathsf{E}(X_{p,t}) \\
                                                     X_{1,t} X_{2,t} & - & \mathsf{E}(X_{1,t} X_{2,t}) \\
                                                     X_{1,t} X_{3,t} & - & \mathsf{E}(X_{1,t} X_{3,t}) \\
                                                        \vdots & & \vdots \\
                                                     X_{p-1,t} X_{p,t} & - & \mathsf{E}(X_{p-1,t} X_{p,t}) \end{pmatrix}
\end{equation*}
and $S_j:= \sum_{t=1}^j U_t$, we have
\begin{align*}
\lim_{m \rightarrow \infty} \mathe\left(\frac{1}{m} S_m
S_m'\right) =: D_1,
\end{align*}
where $D_1$ is a finite and positive definite matrix with $2p+\frac{p(p-1)}{2}$ rows and $2p+\frac{p(p-1)}{2}$ columns.
\end{assumption}

\begin{assumption}\label{moments}
For some $r > 2$, the $r$-th absolute moments of the components of $U_t$ are
uniformly bounded, that means, $\sup_{t \in \mathbb{Z}} \mathsf{E} ||U_t||_{r} < \infty$.
\end{assumption}

\begin{assumption}\label{ned}
For $r$ from Assumption \ref{moments}, the vector $(X_{1,t},\ldots,X_{p,t})$ is $L_2$-NED (near-epoch dependent) with
size $-\frac{r-1}{r-2}$ and constants $(c_{t}),t \in \mathbb{Z}$, on a sequence
$(V_{t}),t \in \mathbb{Z}$, which is $\alpha$-mixing of size $\phi^*
:= -\frac{r}{r-2}$, i.e.,
\begin{align*}
\left|\left|(X_{1,t},\ldots,X_{p,t}) - \mathe\left((X_{1,t},\ldots,X_{p,t}) | \sigma(V_{t-l},\ldots,V_{t+l}) \right)\right|\right|_2 \leq c_t v_l
\end{align*}
with $\lim_{l \rightarrow \infty} v_l = 0$. The constants $(c_{t}),t \in \mathbb{Z}$ fulfill $c_{t} \leq 2 ||U_t||_2$ with $U_t$ from Assumption \ref{limit}.
\end{assumption}

Assumption \ref{limit} is a regularity condition which rules out trending random variables. As we have financial returns in mind, this is no issue.

Assumption \ref{moments} is more critical because it requires finite $|4 + \gamma|$-th moments of ${\bf X}_t$ with an arbitrary $\gamma > 0$ (note that the components of ${\bf X}_t$ enter $U_t$ quadratically).
In fact, there is evidence that even variances might not exist for some financial series, cf. \citet{mandelbrot:1963}.
However, simulation evidence below shows that the test still works under the $t_3$-distribution.

Assumption \ref{ned} is a very general serial dependence assumption which holds in relevant econometric models, e.g.\ in GARCH-models under certain conditions (cf. \citealp{carrascochen:2002}).
It guarantees that the vector
\begin{equation*}
(X^2_{1,t},\ldots,X^2_{p,t},X_{1,t},\ldots,X_{p,t},X_{1,t} X_{2,t},X_{1,t} X_{3,t},\ldots,X_{p-1,t} X_{p,t})
\end{equation*}
is $L_2$-NED (near-epoch dependent) with size $-\frac{1}{2}$, cf. \citet{davidson:1994}, p. 273. This allows for applying a functional central limit theorem later on.

Next, we impose a stationarity condition which is in line with \citet{aue:2009b}.

\begin{assumption}\label{stationarity}
$(X_{1,t},\ldots,X_{p,t}), t \in \mathbb{Z},$ has constant expectation and variances, that means, $\mathsf{E}(X_{i,t}), i = 1,\ldots,p,$ and
$0 < \mathsf{E}(X_{i,t}^2), 1 \leq i \leq p,$ do not depend on $t$.
\end{assumption}

This condition might be
slightly relaxed to allow for some fluctuations in the first and second
moments (see (A4) and (A5) in \citealp{wied:2012}), but for ease of exposition
and because the procedure would remain exactly the same, we stick to this assumption. Note that most financial time series processes as for example GARCH are (unconditionally) stationary under certain conditions.
Clearly, the original test problem is invariant under heteroscedasticity. But we believe that it is at least extremely difficult if not impossible to design a fluctuation test for correlations in which
arbitrary variance changes are allowed under the null hypothesis.

Our main result is:

\begin{theorem}\label{theorem1}
Under $H_0$ and Assumptions \ref{limit},\ref{moments},\ref{ned},\ref{stationarity}, for $T \rightarrow \infty$,
\begin{equation*}
\frac{\tau(s)}{\sqrt{T}} (\hat \rho^{ij}_{\tau(s)} - \hat \rho^{ij}_T)_{1 \leq i < j \leq p} \Rightarrow_d E^{1/2} B^{\frac{p(p-1)}{2}}(s),
\end{equation*}
on $D\left([0,1],\mathbb{R}^{\frac{p(p-1)}{2}}\right)$, where $\tau(s) = \lfloor 2 + s(T-2) \rfloor$, $$E = \lim_{T \rightarrow \infty} \mathsf{Cov} \left( \sqrt{T} \left(\hat \rho^{ij}_T \right)_{1 \leq i < j \leq p} \right) \sim \left(\frac{p(p-1)}{2} \times \frac{p(p-1)}{2}\right)$$ and
$B^{\frac{p(p-1)}{2}}(s)$ is a vector of $\frac{p(p-1)}{2}$ independent standard Brownian Bridges.
\end{theorem}

The proof of the theorem can be found in the appendix. It relies on the application of an adapted functional delta method.
We want to stress that simply applying a functional central limit theorem is not enough here due to the cumbersome, non-linear structure of the correlation coefficient.

From the previous theorem, we directly obtain with the Continuous Mapping Theorem
\begin{cor}\label{corollary1}
Under $H_0$ and Assumptions \ref{limit},\ref{moments},\ref{ned},\ref{stationarity}, for $T \rightarrow \infty$, 
\begin{equation*}
Q_T \rightarrow_d \sup_{0 \leq s \leq 1} \left|\left| E^{1/2} B^{\frac{p(p-1)}{2}}(s) \right|\right|_1.
\end{equation*}
\end{cor}

In order to obtain critical values, we need information about $E$.
There are several possibilities for estimating $E$; one possibility is the estimator $\hat E$, given by a bootstrap approximation. For this estimation, one can for example use the moving block bootstrap from \citet{kuensch:1989} and \citet{liusingh:1992},
cf.\ also \citet{lahiri:1999}, \citet{concalves:2002}, \citet{concalves:2003}, \citet{calhoun:2013}, \citet{radulovic:2012} and \citet{sharipov:2012}.

Defining a block length $l_T$, we divide the time series into $T-l_T-1$ overlapping blocks $B_i, i = 1,\ldots,T-l_t-1,$ with length $l_T$ such that
$B_1 = ({\bf X}_1,\ldots,{\bf X}_{l_T})$, $B_2 = ({\bf X}_2,\ldots,{\bf X}_{l_T+1}),\ldots$.
Then, in each bootstrap repetition $b, b=1,\ldots,B$ for some large $B$, we sample $\left\lfloor\frac{T}{l_T}\right\rfloor$ times with replacement one of the $T-l_T-1$ blocks and concatenate the blocks. So, we obtain $B$ $p$-dimensional time series with length $\left\lfloor\frac{T}{l_T}\right\rfloor \cdot l_T$. For each bootstrapped time series we calculate the vector $v_b := \sqrt{T} \left(\hat \rho^{ij}_{b,T} \right)_{1 \leq i < j \leq p}$.
The estimator $\hat E$ is then the empirical covariance matrix of these $B$ vectors, i.e., $$\hat E = \frac{1}{B} \sum_{b=1}^B (v_b - \bar v)(v_b - \bar v)'$$ with $\bar v = \frac{1}{B} \sum_{b=1}^B v_b$. The bootstrap estimator ``replaces'' the rather complicated kernel estimator $\tilde E$ from the KB-test (Appendix A.1 in \citealp{wied:2012}).
The advantage of the bootstrap estimator is the fact that it can be derived easily even in higher dimensions.
It would be possible to obtain a kernel estimator also in higher $(> 2)$ dimensions. However, its structure would then depend on the structure of derivatives of certain non-linear, higher-dimensional functions which transform a high-dimensional vector of moments to the vector of correlation coefficients. (More information is given in the proof of Theorem \ref{theorem1}). The arguably complicated transformation makes the calculation of a kernel estimator very cumbersome and much harder to implement.
Moreover, a kernel estimator depends on the choice of the bandwidth and the kernel. The disadvantage of the bootstrap is that it is computationally more intensive. In addition, the choice of the block length is required.

The matrix $\hat E$ is an estimator for $\mathsf{Cov^*}(v_b)$ which is the (theoretical) covariance matrix of $v_b$ with respect to the bootstrap sample conditionally on the original data ${\bf X}_1,\ldots,{\bf X}_T$. In order to validate the bootstrap, the key point is the proof that, for $T \rightarrow \infty$, $\mathsf{Cov^*}(v_b)$ converges in probability to $E$. In order to obtain such an asymptotic result, we need an assumption on the block length.

\begin{assumption}\label{bootstrap}
For $T \rightarrow \infty$, $l_T \rightarrow \infty$ and $l_T \sim T^{\alpha}$ for $\alpha \in (0,1)$.
\end{assumption}

The assumption is similar to the one for the moving block bootstrap in Theorem 1 (Condition 4) of \citet{calhoun:2013}. It guarantees that the block length becomes large but not too large compared to $T$.

Moreover, we need an assumption which ensures that the bootstrap correlation coefficients are sufficiently close to the correlation coefficients obtained from the data.

\begin{assumption}\label{uniformintboot}
For $1 \leq i < j \leq p$, some $\delta > 0$ and $b=1,\ldots,B$, the random variable
$$C_T := \mathsf{E}\left(\left|\sqrt{T} (\hat \rho^{ij}_{b,T} - \hat \rho^{ij}_T)\right|^{2+\delta} \left| {\bf X}_1,\ldots,{\bf X}_T \right. \right)$$ is stochastically bounded ($C_T = O_{\mathsf{P}}(1)$).
\end{assumption}

The next theorem gives the theoretical validation for the bootstrap.

\begin{theorem}\label{theorem2}
Under $H_0$ and Assumptions \ref{limit},\ref{moments},\ref{ned},\ref{stationarity},\ref{bootstrap},\ref{uniformintboot}, for $T \rightarrow \infty$, $$\mathsf{Cov^*}(\sqrt{T} \left(\hat \rho^{ij}_{b,T} \right)_{1 \leq i < j \leq p}) \rightarrow_p E.$$
\end{theorem}

Given the theoretical results, it is reasonable to consider the ``test statistic''
\begin{equation*}
A_T := \max_{2 \leq k \leq T} \frac{k}{\sqrt{T}} \left| \left| \hat E^{-1/2} P_{k,T} \right| \right|_1
\end{equation*}
in applications. Then, the null hypothesis is rejected whenever $A_T$ is larger than the $(1-\alpha)$-quantile of the random variable $A := \sup_{0 \leq s \leq 1} \left|\left| B^{\frac{p(p-1)}{2}}(s) \right|\right|_1$. The quantiles of $A$, which serve as an approximation for the quantiles of the finite sample distribution, can easily be obtained by Monte Carlo simulations, i.e., by approximating the paths of the Brownian Bridge on fine grids.

There might be situations in practice in which $\hat E^{1/2}$ is not positive definite so that $\hat E^{-1/2}$ would not be defined. However, due to Assumption \ref{limit}, at least for larger $T$ and $B$, we can virtually assume positive definiteness.\footnote{To circumvent the problem of impossible or numerically unstable inversion of $\hat E^{1/2}$, one could calculate the statistic $Q_T$ and simulate critical values from the limit random variable in Corollary \ref{corollary1} in which $E$ is replaced by $\hat E$.}

\section{Local Power} \label{sec:localpower}

Econometricians are often not only interested in the behavior of a test under the null hypothesis, but would like to get information about the behavior under some local alternatives.
For simplicity, we consider a setting in which the expectations and variances remain constant such that a covariance change is equal to a change in correlations.
To be more precise, under $H_1$, in at least one of the components of ${\bf X}_t$, there is a correlation change of order $\frac{M}{\sqrt{T}}$ ($M > 0$ arbitrary) with constant expectations and variances and
\begin{equation*}
(\mathsf{E}(X_{i,t} X_{j,t}))_{1 \leq i < j \leq p} = v + \frac{M}{\sqrt{T}} g\left( \frac{t}{T} \right).
\end{equation*}
Here, $v \in \mathbb{R}^{\frac{p(p-1)}{2}}$ is a constant vector and $g(s) = (g_1(s),\ldots,g_{\left( \frac{p(p-1)}{2} \right)}(s))$ is a bounded $\frac{p(p-1)}{2}$-dimensional function that is not constant and that can be approximated by step functions such that the function
\begin{equation*}
\int_0^s g(u) du - s \int_0^1 g(u) du
\end{equation*}
is different from $0 \in \mathbb{R}^{\frac{p(p-1)}{2}}$ for at least one $s \in [0,1]$. The integral is defined component by component.

Note that we now deal with triangular arrays because the distribution of the ${\bf X}_t$ changes with $T$, but, for simplicity, we do not change our notation.

A typical example for the function $g$ would be a step function with a jump from $0$ to $g_0$ in a given point $z_0$ in one of the components. This implies that the correlation of one pair jumps at time $\lfloor T \cdot z_0 \rfloor$. A step function with several jumps would correspond to multiple change points. With a continuous function $g$, one would obtain continuously changing correlations.

The following Theorem \ref{theorem3} is an analogue to Theorem \ref{theorem1} and yields the distribution under the sequence of local alternatives.
\begin{theorem}\label{theorem3}
Under the sequence of local alternatives and Assumptions \ref{limit},\ref{moments},\ref{ned},\ref{stationarity}, for $T \rightarrow \infty$,
\begin{equation*}
\frac{\tau(s)}{\sqrt{T}} (\hat \rho^{ij}_{\tau(s)} - \hat \rho^{ij}_T)_{1 \leq i < j \leq p} \Rightarrow_d E^{1/2} B^{\frac{p(p-1)}{2}}(s) + E^{1/2} C(s),
\end{equation*}
on $D\left([0,1],\mathbb{R}^{\frac{p(p-1)}{2}}\right)$, where
\begin{equation*}
C(s) = M \begin{pmatrix} \frac{1}{\sqrt{\mathsf{Var}(X_{1})\mathsf{Var}(X_{2})}} \left( \int_0^s g_1(u)du - s \int_0^1 g_1(u) du \right) \\ \frac{1}{\sqrt{\mathsf{Var}(X_{1})\mathsf{Var}(X_{3})}} \left( \int_0^s g_2(u)du - s \int_0^1 g_2(u) du \right) \\ \vdots \\ \frac{1}{\sqrt{\mathsf{Var}(X_{p-1})\mathsf{Var}(X_{p})}} \left( \int_0^s g_{\frac{p(p-1)}{2}}(u)du - s \int_0^1 g_{\frac{p(p-1)}{2}}(u)du \right) \end{pmatrix}
\end{equation*}
is a deterministic function that depends on the specific form of the local alternative under consideration, characterized by $g$.
\end{theorem}
In Theorem \ref{theorem3}, the supremum is taken over the absolute value of a Brownian Bridge plus a
deterministic function $C(s)$.
As the main characteristic of the function $C(s)$ from Theorem \ref{theorem3}, we have the factor $M$ times the expression $\int_0^s g_i(u)du - s \int_0^1 g_i(u)du$ in each component $i=1,\ldots, \frac{p(p-1)}{2}$. This follows from the structure of a Brownian Bridge.

The previous theorem directly yields with the Continuous Mapping Theorem
\begin{cor}
Under the sequence of local alternatives and Assumptions \ref{limit},\ref{moments},\ref{ned},\ref{stationarity}, for $T \rightarrow \infty$,
\begin{equation*}
Q_T \rightarrow_d \sup_{0 \leq s \leq 1} \left| \left| E^{1/2} B^{\frac{p(p-1)}{2}}(s) + E^{1/2} C(s)  \right| \right|_1
\end{equation*}
\end{cor}

Also under local alternatives we want to estimate $E$ with the bootstrap. It turns out that the estimator presented in Section \ref{sec:retrotest} has the same limit distribution as under the null hypothesis.
Thus, the bootstrap approach is valid both under the null and under the alternative.

\begin{theorem}\label{theorem4}
Under the sequence of local alternatives and Assumptions \ref{limit},\ref{moments},\ref{ned},\ref{stationarity},\ref{bootstrap},\ref{uniformintboot}, for $T \rightarrow \infty$, $$\mathsf{Cov^*}(\sqrt{T} \left(\hat \rho^{ij}_{b,T} \right)_{1 \leq i < j \leq p}) \rightarrow_p E.$$
\end{theorem}

The theoretical results in this section imply that the quantity $A_T$ is close to $A_L := \sup_{0 \leq s \leq 1} \left| \left| B^{\frac{p(p-1)}{2}}(s) + C(s)  \right| \right|_1$ for large $T$ and $B$. Moreover, for every $B \geq 1$, the test statistic becomes arbitrarily large for large $M$ and $T$.

\section{Finite Sample Evidence} \label{sec:simulations}
We illustrate the finite sample properties of our multivariate test with Monte Carlo simulations in different settings:
We consider a series of four-variate random vectors which are, on the one hand, serially independent and, on the other hand, fulfill a four-variate MA(1)-structure with MA-parameters $0.5$. This means that, for $t=0,\ldots,T$, there are serially independent vectors ${\bf u}_t := (u_{t,1},u_{t,2},u_{t,3},u_{t,4})$ such that the data generating process is defined by $${\bf X}_t = {\bf u}_t + A {\bf u}_{t-1}, t=1,\ldots,T$$ with ${\bf X}_t = (X_{t,1},X_{t,2},X_{t,3},X_{t,4}), A=\text{diag}(\theta,\theta,\theta,\theta)$ and $\theta \in \{0,0.5\}$.
The lengths of the series are chosen as $T \in \{200,500\}$, the block lengths are $l_T = \lfloor T^{1/4} \rfloor$, respectively\footnote{For, $T=200$, $\left\lfloor\frac{T}{l_T}\right\rfloor \neq \frac{T}{l_T}$, so that the length of the bootstrapped time series is not exactly equal to $T$.
However, we consider the difference as negligible.}, the number of bootstrap replications is $999$ and the number of Monte Carlo replications is
$10000$. We consider, on the one hand, a four-variate normal distribution (ND) and, on the other hand, a four-variate $t_{3}$-distribution.
The $t_3$-distribution is not covered by our assumptions, but we analyze it to get a picture of the behavior of the test in settings which are realistic in financial applications.

For simulating the behavior under the null, we set the variances of the $u_{i,t}, i=1,2,3,4,$ to $1$ and the correlations of the $u_{i,t}$ to $\rho_{12} = \ldots = \rho_{34} =: \rho_0 \in \{0,0.5\}$. Under the alternative, the $u_{i,t}$ have correlation $\rho_0$ in the first half of the sample. Moreover, we have a change in all six pairwise correlations of the $u_{i,t}$ with shifts $\Delta \rho = -0.2,-0.4,0.2,0.4$ in the middle of the sample.

The results (empirical rejection probabilities, not-size-adjusted, nominal level $0.05$ which corresponds to a simulated critical value of $4.47$) are given in Table \ref{table1}.

\begin{center}
- Table \ref{table1} here -
\end{center}

It is seen that there are some size distortions for the heavy-tailed distribution and/or serial dependence although the level seems to converge to $0.05$ for higher $T$ in all cases.
The power of the test increases in $T$ and in absolute values of the correlation changes.
For the $t_3$-distribution, the power is in general considerably lower. Further simulations show that the size properties become worse for even higher $\rho_0$ and even higher serial dependence.

Moreover, we compare the test for constant correlation matrix with a multivariate procedure based on the pairwise correlation test from \citet{wied:2012} (with bandwidth $\lfloor \log(T) \rfloor$) and the Bonferroni-Holm correction. That means that we perform $m=6$ pairwise tests and denote by $p_{(1)},\ldots,p_{(m)}$ the corresponding p-values in increasing order. We declare the null hypothesis of constant correlation matrix to be invalid if there is at least one $j=1,\ldots,m$ such that $$p_{(j)} < \frac{0.05}{m+1-j}.$$ The results are also presented in Table \ref{table1}. Depending on the situation, sometimes the one and sometimes the other procedure performs better. While the Bonferroni-Holm procedure has in general slightly better size and power properties for $\rho_0=0.5$, the matrix-based test performs better with $\rho_0=0$, especially with the normal distribution and decreasing correlation. There is even one case in which the Bonferroni-Holm procedure is not unbiased (the power is smaller than the size) which does not occur with the matrix-based test.

In another setting, we have compared the bivariate bootstrap with the bivariate kernel-based and have seen that both tests behave more or less similarly.

\section{Application to Stock Returns}
\begin{figure}[h]
\subfigure[Rolling correlations between Total and Sanofi]{\includegraphics[scale=0.3]{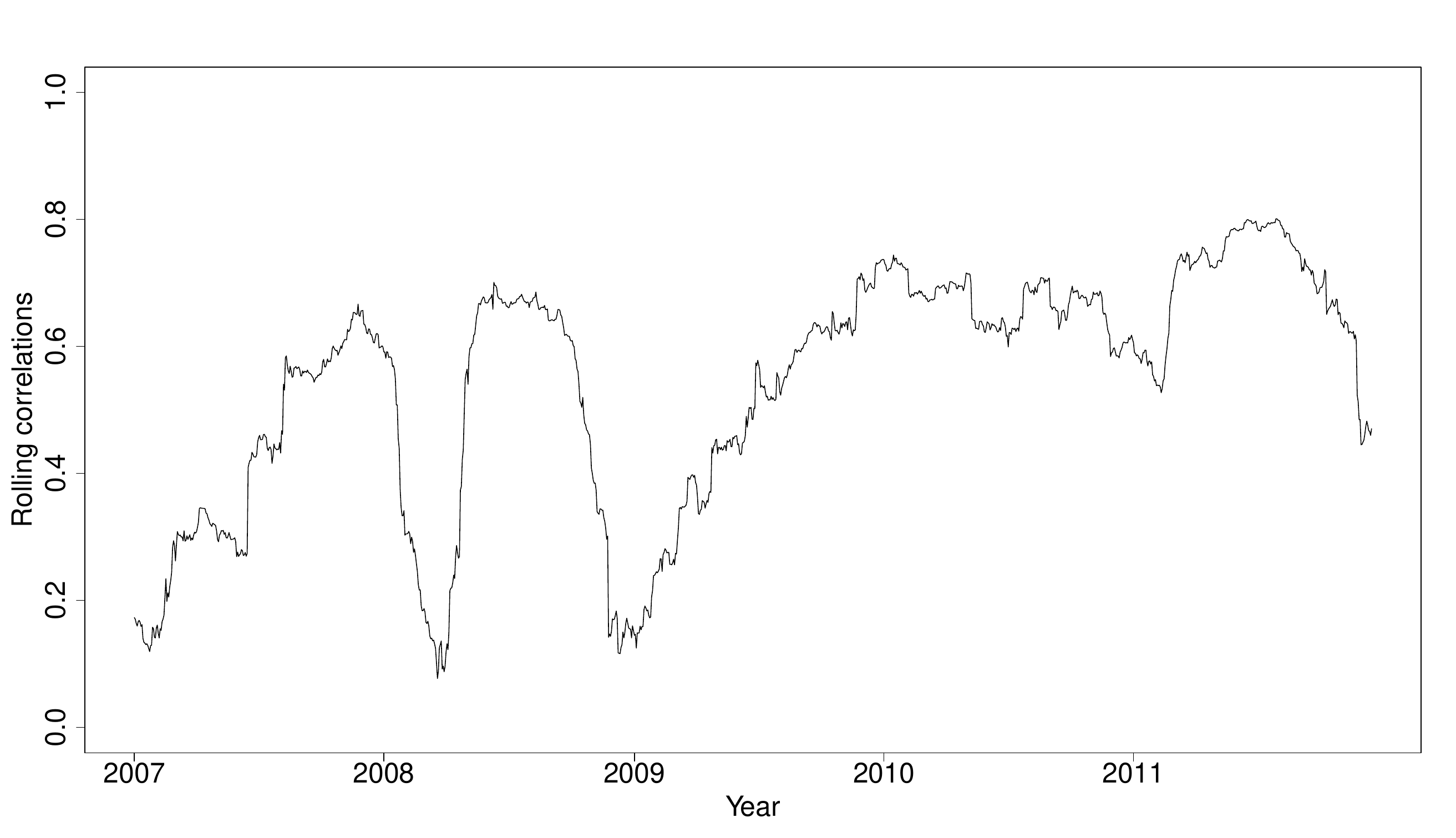}}\hfill
\subfigure[Rolling correlations between Total and Siemens]{\includegraphics[scale=0.3]{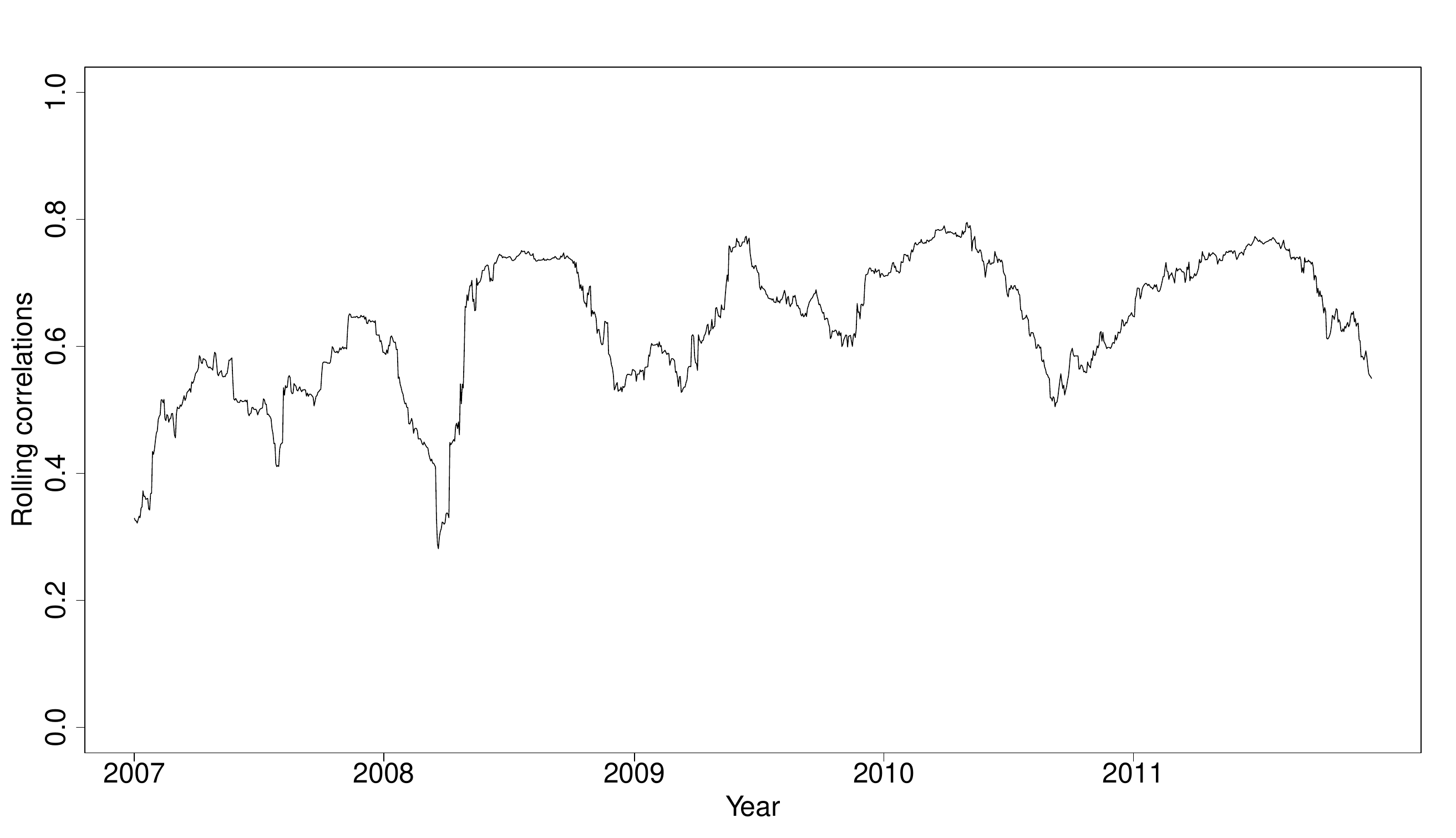}}
\caption{Rolling correlations}\label{fig:correlations1}
\end{figure}

\begin{figure}[h]
\subfigure[Rolling correlations between Total and BASF]{\includegraphics[scale=0.3]{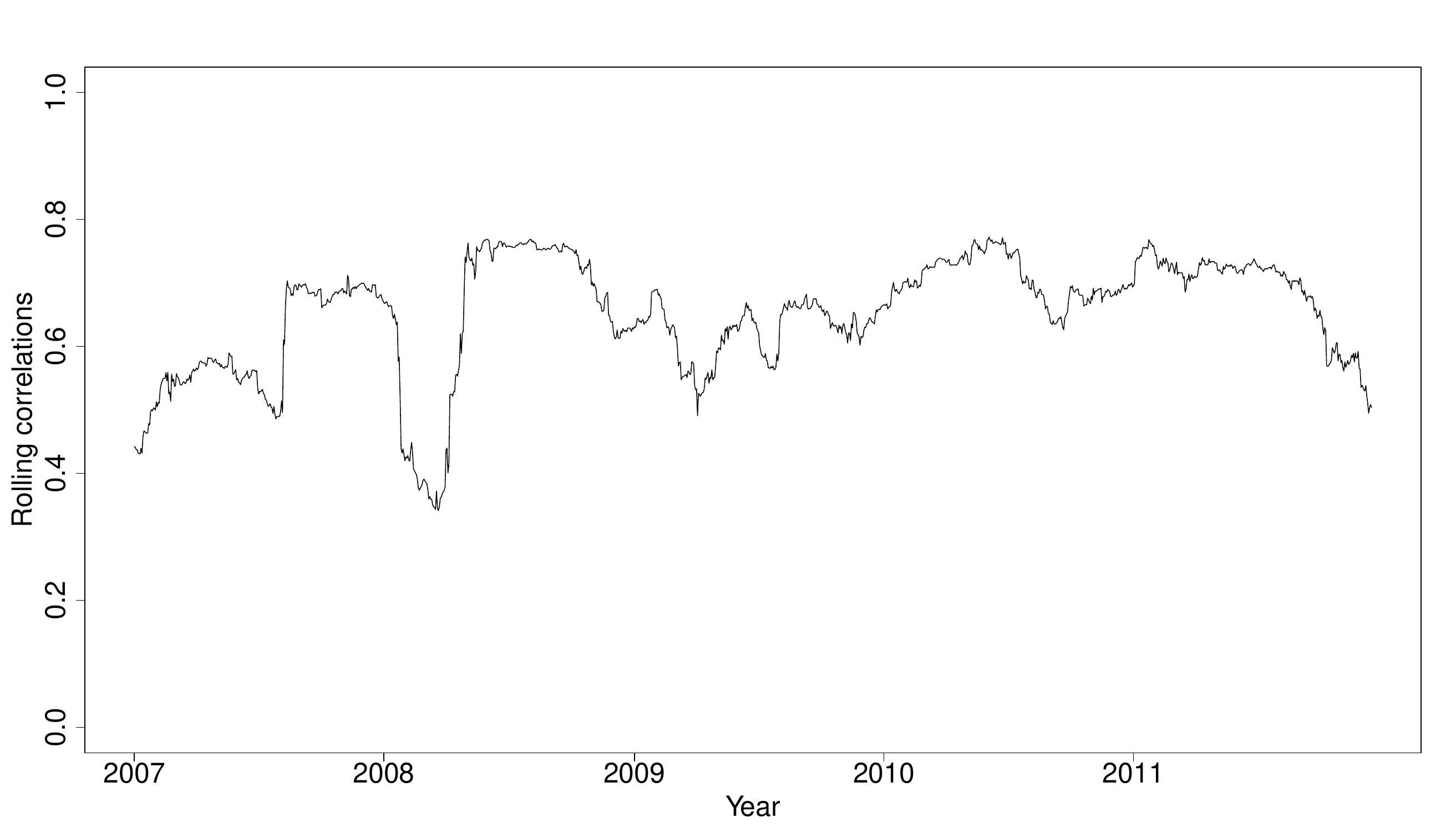}}\hfill
\subfigure[Rolling correlations between Sanofi and Siemens]{\includegraphics[scale=0.3]{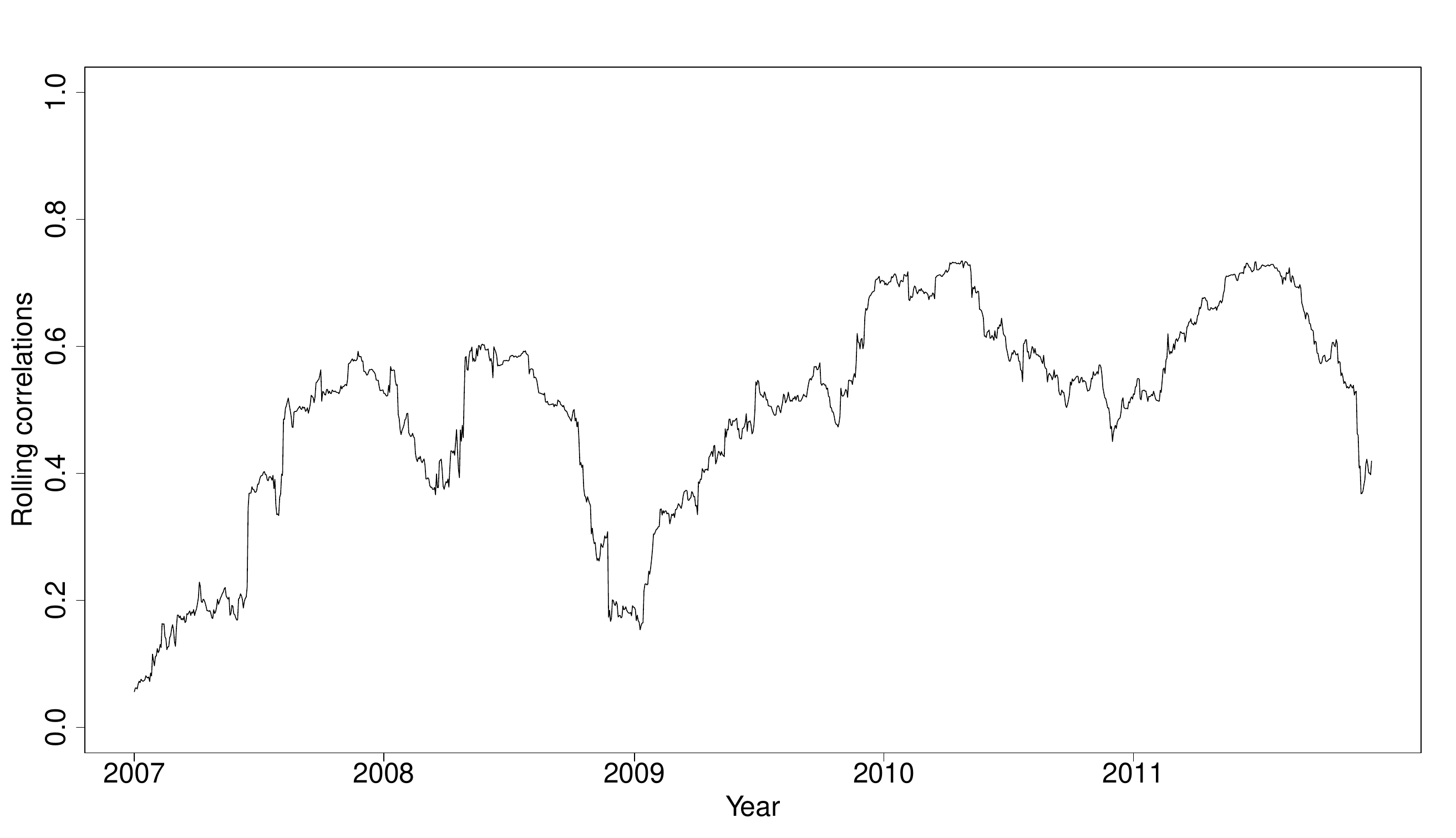}}
\caption{Rolling correlations}\label{fig:correlations2}
\end{figure}

\begin{figure}[h]
\subfigure[Rolling correlations between Sanofi and BASF]{\includegraphics[scale=0.3]{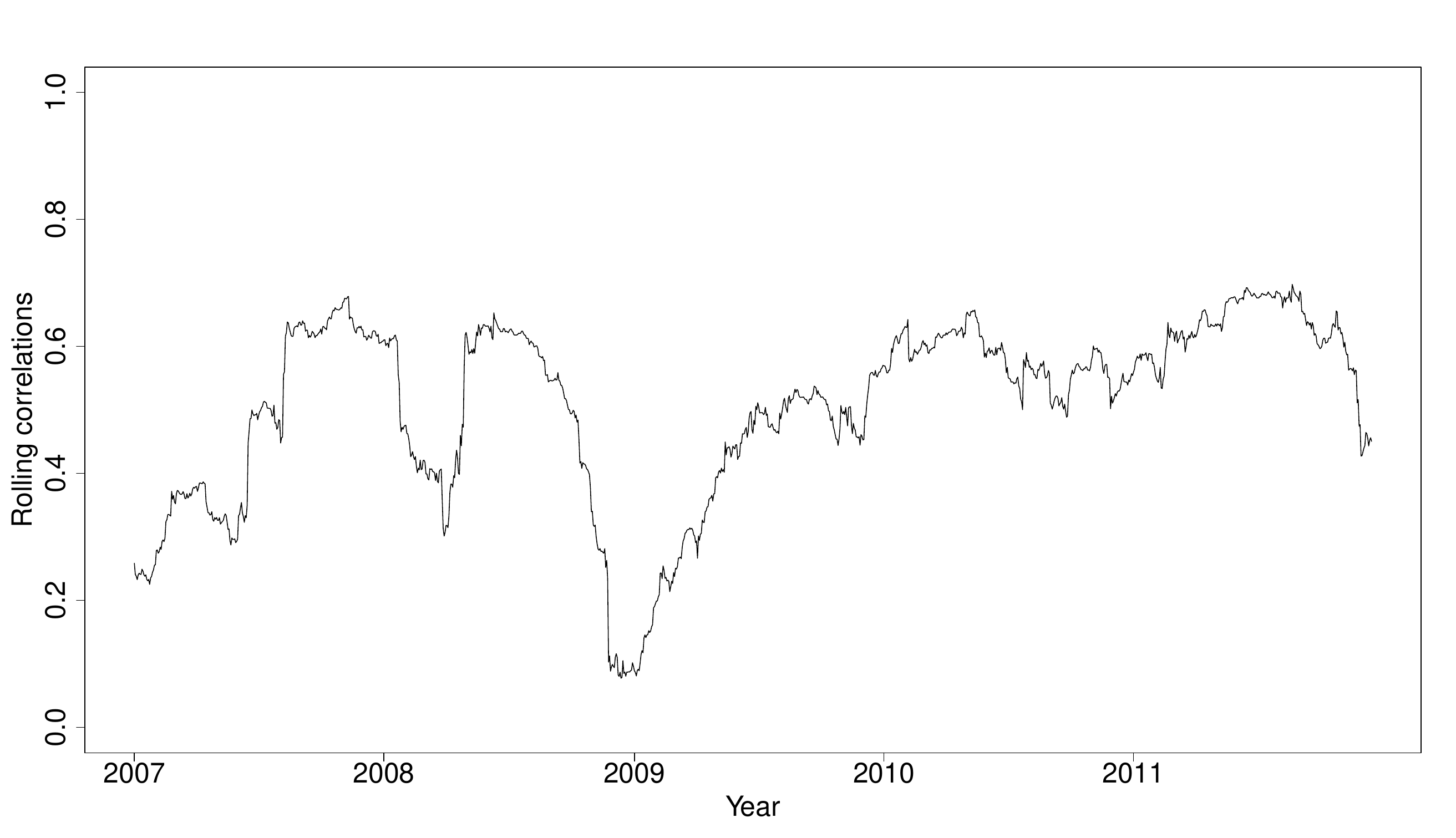}}\hfill
\subfigure[Rolling correlations between Siemens and BASF]{\includegraphics[scale=0.3]{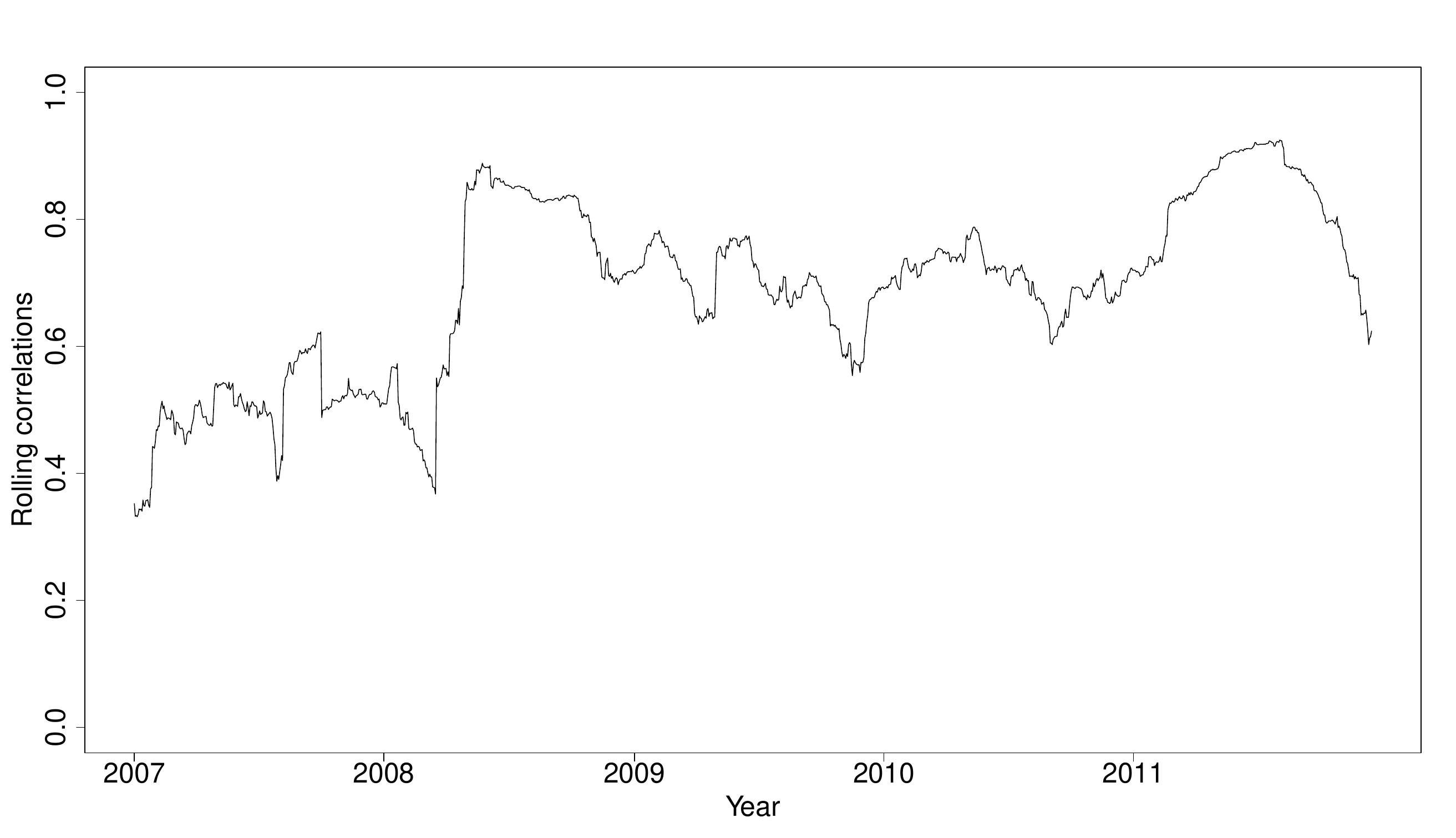}}
\caption{Rolling correlations}\label{fig:correlations3}
\end{figure}

Next, we show how the proposed test can be applied in financial time series. For this, we consider the correlation of four stocks.
In order to avoid issues due to market trading in different time zones, we just consider the European market.
We look at the four companies of Euro Stoxx 50 with the highest weights in the index in the end of May 2012, that means Total, Sanofi, Siemens and BASF,
and consider the time span 01.01.2007 - 01.06.2012 such that $T=1414$.
The data was obtained from the database Datastream. Figures \ref{fig:correlations1}, \ref{fig:correlations2}, \ref{fig:correlations3} plot rolling windows of the six pairwise correlations
of the continuous daily returns from each asset with the window length $120$. This corresponds to the trading time of about half a year.
The days on the x-axis show the first day of the windows, respectively.

\begin{center}
- Figure \ref{fig:correlations1} here -

- Figure \ref{fig:correlations2} here -

- Figure \ref{fig:correlations3} here -
\end{center}

We identify time-varying correlations. It is for example interesting to see that the correlation between Total and Sanofi is close to
$0$ in the beginning of February 2008 and much higher after this. The correlation between Sanofi and BASF is interestingly low in the middle of 2009.

The test statistic $Q_T$ applied on the four-variate return vector is equal to $10.49$. With $B=10999$ bootstrap replications, we obtain $A_T = 6.55$. With this value, we cannot yet determine if the null hypothesis of constant correlation is rejected.
So we calculate the statistic $A_T$ with $B=10999$. The $0.95$-quantile of $\sup_{0 \leq s \leq 1} \left| \left| B^{6}(t) \right| \right|_1$ is equal to $4.47$ and so, the null hypothesis is rejected on the
significance level $\alpha=0.05$. The approximate p-value is smaller than $0.001$.

\begin{figure}[h]
\includegraphics[scale=0.3]{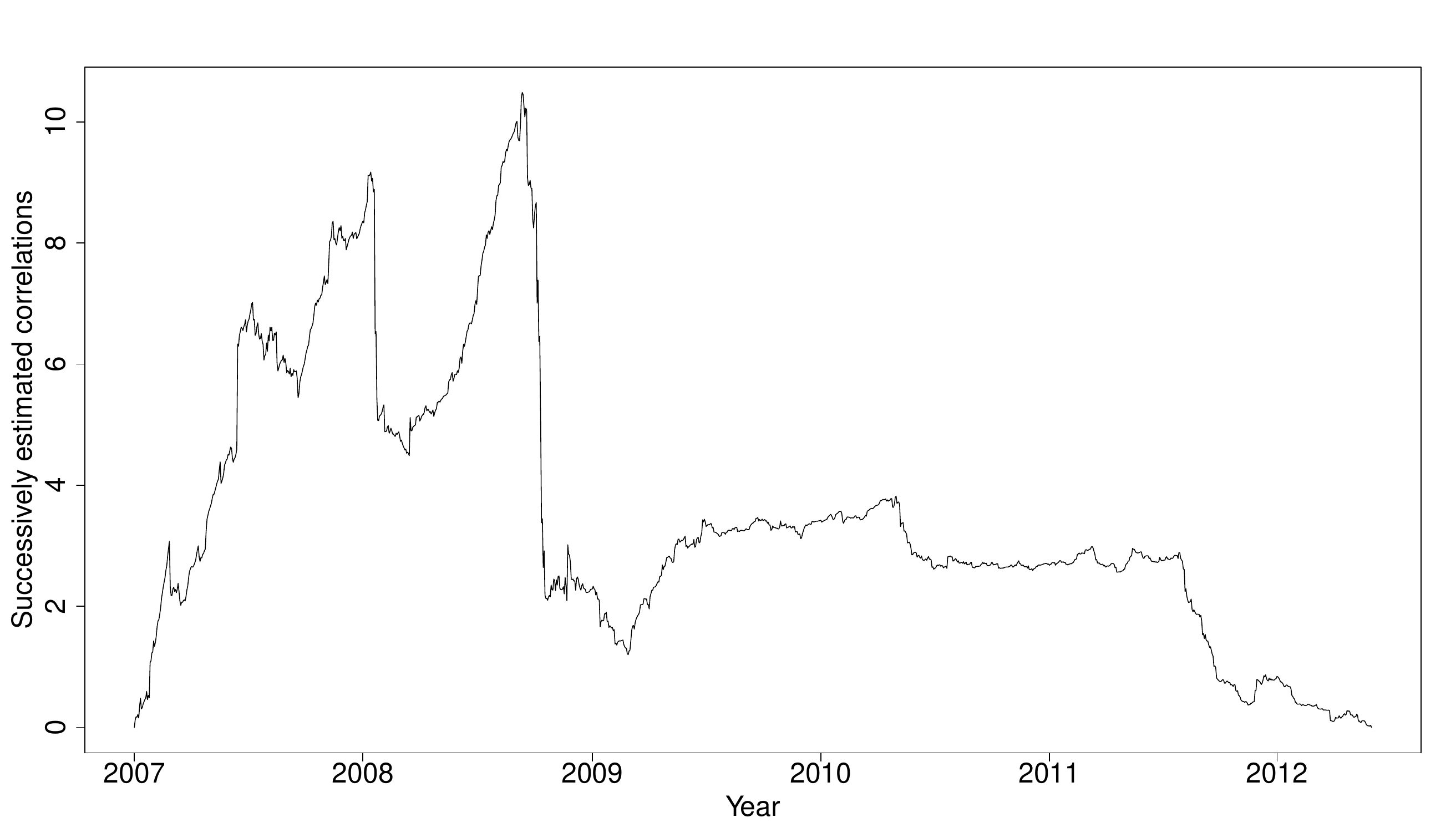}
\caption{Evolution of successively calculated correlations}\label{fig:evolution}
\end{figure}

Figure \ref{fig:evolution} shows the process
\begin{equation*}
\left( \sum_{1 \leq i < j \leq p} \frac{k}{\sqrt{T}} \left| \hat \rho^{ij}_k - \hat \rho^{ij}_T \right| \right)_{2 \leq k \leq T},
\end{equation*}
that means the evolution of the successively calculated correlations over time.
In the context of CUSUM tests, the point of the maximum is often considered as a reasonable estimator for the (most important) change point if the test decides that such a point actually exists,
see \citet{vostrikova:1981} and the related literature. In this case, the maximum is obtained at the 11th of September 2008 which corresponds quite well to the insolvency of Lehman Brothers. A discussion on dating multiple change points in the correlation matrix can be found in \citet{galeano:2014}.

\begin{center}
- Figure \ref{fig:evolution} here -
\end{center}

\section{Conclusion}
We have presented a new fluctuation test for constant correlations in the multivariate setting for which the location of potential change points need not be specified a priori. The new test is based on a bootstrap approximation, works under mild assumptions regarding the dependence structure, has appealing properties in simulations and seems to be useful in empirical applications. Potential drawbacks of the test are the requirement of finite fourth moments and the assumption of constant expectations and variances. It might be an interesting question for the future to thoroughly investigate to which extent these drawbacks could be overcome by some kind of prefiltering and/or other transformations. Moreover, it could be worthwile to extend the present approach to the problem of monitoring correlation changes or to other, perhaps more robust measures of dependence.

\newpage

\bibliographystyle{ecta}
\bibliography{wied_multivariate_correlation_testing}
\newpage

\appendix

\section{Appendix}
{\it Proof of Theorem \ref{theorem1}} \\
Note that the null hypothesis and Assumption \ref{stationarity} imply that $\mathsf{E}(X_{i,t} X_{j,t}), 1 \leq i,j \leq p,$ do not depend on $t$.

At first, we need an invariance principle for the vector
\begin{equation*}
V_T(s) := \frac{1}{\sqrt{T}} \sum_{t=1}^{\tau(s)} \begin{pmatrix} X^2_{1,t} & - & \mathsf{E}(X^2_{1,t}) \\
                                                        \vdots & & \vdots \\
                                                     X^2_{p,t} & - & \mathsf{E}(X^2_{p,t}) \\
                                                     X_{1,t} & - & \mathsf{E}(X_{1,t}) \\
                                                        \vdots & & \vdots \\
                                                     X_{p,t} & - & \mathsf{E}(X_{p,t}) \\
                                                     X_{1,t} X_{2,t} & - & \mathsf{E}(X_{1,t} X_{2,t}) \\
                                                     X_{1,t} X_{3,t} & - & \mathsf{E}(X_{1,t} X_{3,t}) \\
                                                        \vdots & & \vdots \\
                                                     X_{p-1,t} X_{p,t} & - & \mathsf{E}(X_{p-1,t} X_{p,t}) \end{pmatrix},
\end{equation*}
which is provided by \citet{davidson:1994}, p. 492. Thus, it holds, for $T \rightarrow \infty$, $V_T(s) \Rightarrow_d D_1^{1/2} W^{2p+\frac{p(p-1)}{2}}(s)$ on $D\left([0,1],\mathbb{R}^{2p+\frac{p(p-1)}{2}}\right)$, where $W^{2p+\frac{p(p-1)}{2}}(s)$ is a $\left(\frac{p(p-1)}{2}+2p\right)$-dimensional Brownian Motion with independent components and $D_1$ is given in Assumption \ref{limit}.

Now, one makes the observation that
\begin{equation*}
V_T(s) = \frac{\tau(s)}{\sqrt{T}} \begin{pmatrix} \overline{X^2_{1}}(s) & - & \mathsf{E}(X^2_{1,t}) \\
                                                        \vdots & & \vdots \\
                                                   \overline{X^2_{p}}(s) & - & \mathsf{E}(X^2_{p,t}) \\
                                                   \overline{X_{1}}(s) & - & \mathsf{E}(X_{1,t}) \\
                                                        \vdots & & \vdots \\
                                                   \overline{X_{p}}(s) & - & \mathsf{E}(X_{p,t}) \\
                                                   \overline{X_{1} X_{2}}(s) & - & \mathsf{E}(X_{1,t} X_{2,t}) \\
                                                   \overline{X_{1} X_{3}}(s) & - & \mathsf{E}(X_{1,t} X_{3,t}) \\
                                                        \vdots & & \vdots \\
                                                   \overline{X_{p-1} X_{p}}(s) & - & \mathsf{E}(X_{p-1,t} X_{p,t}) \end{pmatrix},
\end{equation*}
where, for $i=1,\ldots,p$, $\overline{X^2_{i}}(s) = \frac{1}{\tau(s)} \sum_{t=1}^{\tau(s)} X_{i,t}$, $\overline{X^2_{i}}(s) = \frac{1}{\tau(s)} \sum_{t=1}^{\tau(s)} X^2_{i,t}$ and, for $1 \leq i < j \leq p$, $\overline{X_{i} X_{j}}(s) = \frac{1}{\tau(s)} \sum_{t=1}^{\tau(s)} X_{i,t} X_{j,t}$. The goal is to transform this vector of simple first and second order moments into the vector with the successively calculated correlation coefficients and then to apply the adapted functional delta method, Theorem A.1 in \citet{wied:2012}.
The transforming functions are
\begin{equation*}
\begin{split}
f_1: \mathbb{R}^{2p+\frac{p(p-1)}{2}} &\rightarrow \mathbb{R}^{p+\frac{p(p-1)}{2}} \\
     (x_1,\ldots,x_{\left(2p+\frac{p(p-1)}{2}\right)}) &\rightarrow \begin{pmatrix} x_1 & - & (x_{p+1}^2) \\ \vdots & & \vdots \\ x_p & - & (x_{2p}^2) \\ x_{2p+1} & - & x_p x_{p+1} \\ x_{2p+2} & - & x_p x_{p+2} \\ \vdots & & \vdots \\ x_{\left(2p + \frac{p(p-1)}{2}\right)} & - & x_{2p-1} x_{2p} \end{pmatrix}
\end{split}
\end{equation*}
for the transformation on the vector of variances and covariances and
\begin{equation*}
\begin{split}
f_2: \mathbb{R}^{p+\frac{p(p-1)}{2}} &\rightarrow \mathbb{R}^{\frac{p(p-1)}{2}} \\
     (x_1,\ldots,x_{\left(p+\frac{p(p-1)}{2}\right)}) &\rightarrow \begin{pmatrix} \frac{x_{p+1}}{\sqrt{x_1 x_2}} \\ \frac{x_{p+2}}{\sqrt{x_1 x_3}} \\ \vdots \\ \frac{x_{\left(p+\frac{p(p-1)}{2}\right)}}{\sqrt{x_{p-1} x_p}} \end{pmatrix}
\end{split}
\end{equation*}
for the transformation on the vector of correlations.

We obtain, for $T \rightarrow \infty$ and for arbitrary $\epsilon > 0$,

\begin{equation}\label{analogueLemma3}
W_T(s) := \frac{\tau(s)}{\sqrt{T}} (\hat \rho^{ij}_{\tau(s)} - \rho^{ij})_{1 \leq i < j \leq p} \Rightarrow_d D_3 D_2 D_1^{1/2} W^{2p+\frac{p(p-1)}{2}}(s)
\end{equation}
on $D\left([\epsilon,1],\mathbb{R}^{2p+\frac{p(p-1)}{2}}\right)$ for matrices $D_2 \sim \left( \left(p+\frac{p(p+1)}{2}\right) \times \left(2p+\frac{p(p+1)}{2}\right)\right)$ and $D_3 \sim \left( \frac{p(p+1)}{2} \times \left(p+\frac{p(p+1)}{2}\right)\right)$. Here, $D_2$ is the Jacobian matrix of $f_1$ and $D_3$ is the Jacobian matrix of $f_2$, evaluated at certain moments.

We are not interested in the exact (and cumbersome) structure of these matrices. But we observe that $D_2$ contains all $\left(p+\frac{p(p+1)}{2}\right)$-dimensional unit vectors and $D_3$ contains all $\left(\frac{p(p+1)}{2}\right)$-dimensional unit vectors (weighted with some constants) in its columns. Thus, $D_2$ and $D_3$ have full column rank. Together with Assumption \ref{limit}, this implies that $D_3 D_2 D_1^{1/2}$ has full column rank. Consequently, $D_3 D_2 D_1 D_2 ' D_3 '$ is invertible and positive definite.

Now, with an application of Theorem 4.2 in \citet{billingsley:1968}, we obtain, for $T \rightarrow \infty$, $W_T(s) \Rightarrow_d D_3 D_2 D_1^{1/2} W^{2p+\frac{p(p-1)}{2}}(s)$ on $D\left([0,1],\mathbb{R}^{2p+\frac{p(p-1)}{2}}\right)$. Moreover, it holds
\begin{equation*}
D_3 D_2 D_1^{1/2} W^{2p+\frac{p(p-1)}{2}}(s) \stackrel{d}{=} (D_3 D_2 D_1 D_2 ' D_3 ')^{1/2} W^{\frac{p(p-1)}{2}}(s)
\end{equation*}
and from \eqref{analogueLemma3} it is easy to see (with $s=1$) that the asymptotic covariance matrix of $\sqrt{T} \left(\hat \rho^{ij}_T \right)_{1 \leq i < j \leq p}$ is equal to
$D_3 D_2 D_1 D_2 ' D_3 ' =: E$. \hfill $\blacksquare$ \\

{\it Proof of Theorem \ref{theorem2}} \\
We use a bootstrap theorem for near epoch dependent data for $V_T(1)$. Note that a univariate bootstrap central limit theorem conditionally on the original data
(see for example \citealp{pauly:2009}, Lemma and Definition 2.7, for a precise definition of this type of convergence) is obtained by \citet{calhoun:2013}, Corollary 2.

For the multivariate generalization, we use an argument based on the Cramér-Wold device.
Since we consider convergence of conditional distributions which are random variables and since an uncountable union of null sets is not necessary a null set again,
we cannot directly apply the Cramér-Wold device. However, we can use an argument based on the Cramér-Wold device and Assumption \ref{limit} for the multivariate
generalization (see \citealp{pauly:2009}, Theorem 3.19, Theorem 3.20 and the related material in this reference; the main argument is that we just consider rational $\lambda$ when applying the Cramér-Wold device).

Then, Condition 1 of \citet{calhoun:2013}, Corollary 2, is fulfilled with our Assumption \ref{ned},
Condition 2 as well as the condition ``$\sum_{t=1}^n (\mu_{nt} - \bar \mu)^2 = o(n^{1/2})$'' with our Assumption \ref{limit} and our Assumption \ref{stationarity}, Condition 3 with our Assumption \ref{moments} and Condition 4 with our Assumption \ref{bootstrap}.

Summing up the previous discussion, the block bootstrap consistently estimates the distribution law of $V_T(1)$.
But then, with the standard (functional) delta method for the bootstrap (\citealp{vandervaart:1996}, Theorem 3.9.11.) transforming $V_T(1)$ to $W_T(1)$,
also the law of $W_T(1)$ is consistently estimated. That means that, for $T \rightarrow \infty$, $$d\left(\mathcal{L}\left(\sqrt{T} (\hat \rho^{ij}_{b,T} - \hat \rho^{ij}_T)_{1 \leq i < j \leq p} | {\bf X}_1,\ldots,{\bf X}_T\right),\mathcal{L} \left(D_3 D_2 D_1^{1/2} W^{2p+\frac{p(p-1)}{2}}(1)\right)\right) \rightarrow_p 0,$$ where $d$ is a metric of weak convergence (see \citealp{pauly:2009}, p. 36) and $\mathcal{L}(\cdot)$ denotes the distribution of a random vector.

Now consider, for $1 \leq i < j \leq p$ and the $\delta$ from Assumption \ref{uniformintboot}, the conditional expectation $$\mathsf{E}\left(\left|\sqrt{T} (\hat \rho^{ij}_{b,T} - \hat \rho^{ij}_T)\right|^{2+\delta} \left| {\bf X}_1,\ldots,{\bf X}_T \right. \right) =: C_T.$$ By Assumption \ref{uniformintboot}, $C_T$ is stochastically bounded. Then, with Lemma 1 in \citealp{cheng:2011}, we can consistently estimate the asymptotic covariance matrix of $W_T(1)$. \hfill $\blacksquare$ \\
{\it Proof of Theorem \ref{theorem3}} \\
Transferring the proof of Theorem \ref{theorem1}, we obtain, under $H_1$, for $T \rightarrow \infty$, $V_T(s) \Rightarrow_d D_1^{1/2} W^{2p+\frac{p(p-1)}{2}}(s) + A(s)$ on $D\left([0,1],\mathbb{R}^{2p+\frac{p(p-1)}{2}}\right)$. Here,
\begin{equation*}
A = \left( 0 , \ldots , 0 , \int_0^s g(u)' du \right)'
\end{equation*}
(note that $g(u)'$ is the transpose of the function $g$).

So,
\begin{equation}\label{analogueLemma3LocalPower}
W_T(s) := \frac{\tau(s)}{\sqrt{T}} (\hat \rho^{ij}_{\tau(s)} - \rho^{ij})_{1 \leq i < j \leq p} \Rightarrow_d D^{1/2} W^{\frac{p(p-1)}{2}}(s) + D_3 D_2 A(s),
\end{equation}
where $D_3$ and $D_2$ are the matrices mentioned in the proof of Theorem \ref{theorem1}. Due to the structure of $D_3$ and $D_2$, we have

\begin{equation*}
D_3 D_2 A(s) = M \begin{pmatrix} \frac{1}{\sqrt{\mathsf{Var}(X_{1})\mathsf{Var}(X_{2})}} \int_0^s g_1(u)du \\ \frac{1}{\sqrt{\mathsf{Var}(X_{1})\mathsf{Var}(X_{3})}} \int_0^s g_2(u)du \\ \vdots \\ \frac{1}{\sqrt{\mathsf{Var}(X_{p-1})\mathsf{Var}(X_{p})}} \int_0^s g_{\frac{p(p-1)}{2}}(u)du \end{pmatrix}.
\end{equation*}
This completes the proof. \hfill $\blacksquare$ \\

{\it Proof of Theorem \ref{theorem4}} \\
Under local alternatives, for $i=1,\ldots,p$, $\mathsf{E}(X_{i,t})$ and $\mathsf{E}(X^2_{i,t})$ are constant, respectively. Moreover, for $1 \leq i < j \leq p$, it holds 
\begin{equation*}
\sum_{t=1}^T \left(\mathsf{E}(X_{i,t} X_{j,t}) - \overline{X_{i} X_{j}}(1)\right)^2 = o(T^{1/2}).
\end{equation*}
Therefore, the condition ``$\sum_{t=1}^n (\mu_{nt} - \bar \mu)^2 = o(n^{1/2})$'' of Corollary 2 in \citet{calhoun:2013} is fulfilled. The other conditions are fulfilled with the same arguments as in Theorem \ref{theorem2}. Then, by this corollary and the Cramér-Wold Theorem, we estimate $E$ consistently with the bootstrap estimator as described in the proof of Theorem \ref{theorem2}. \hfill $\blacksquare$ \\

\begin{table}
\centering
\caption{Empirical size and empirical power (times $100$, respectively) of the multivariate correlation test; columns 5,6 give empirical rejection probabilities for the matrix-based test, columns 7,8 give rejection probabilites for the Bonferroni-Holm procedure}
\vspace{2.0ex}
\begin{tabular}{|c|c|c|cc|cc|}\hline
$MA$ & distr. & $\rho_0$ & rej.prob. &         & rej.prob. & \\ \hline
     &        &          & $T=200$   & $T=500$ & $T=200$   & $T=500$ \\ \hline \hline
		\multicolumn{7}{|c|}{$\Delta \rho=0$} \\ \hline
		0 & N & 0 & 2.8 & 3.8 & 2.9 & 3.5 \\
		0 & N & 0.5 & 4.4 & 4.4 & 5.5 & 4.3 \\
		0 & t & 0 & 8.7 & 6.7 & 7.8 & 4.9 \\
		0 & t & 0.5 & 11.5 & 8.1 & 10.1 & 6.3 \\ \hline
		0.5 & N & 0 & 4.8 & 5.3 & 4.0 & 4.0 \\
		0.5 & N & 0.5 & 7.4 & 6.2 & 7.5 & 5.1 \\
		0.5 & t & 0 & 13.1 & 9.4 & 9.7 & 5.5 \\
		0.5 & t & 0.5 & 17.1 & 12.3 & 13.2 & 8.1 \\ \hline \hline
		\multicolumn{7}{|c|}{$\Delta \rho=0.2$} \\ \hline
		0 & N & 0 & 32.2 & 89.1 & 26.3 & 73.5 \\
		0 & N & 0.5 & 43.6 & 90.5 & 55.6 & 93.7 \\
		0 & t & 0 & 19.0 & 32.3 & 15.8 & 20.2 \\
		0 & t & 0.5 & 30.1 & 41.3 & 35.4 & 44.9 \\ \hline
		0.5 & N & 0 & 30.3 & 80.8 & 24.7 & 63.9 \\
		0.5 & N & 0.5 & 42.3 & 82.9 & 52.3 & 87.9 \\
		0.5 & t & 0 & 24.5 & 34.7 & 18.4 & 22.2 \\
		0.5 & t & 0.5 & 36.6 & 44.8 & 38.9 & 46.7 \\ \hline \hline
		\multicolumn{7}{|c|}{$\Delta \rho=-0.2$} \\ \hline
		0 & N & 0 & 74.4 & 100.0 & 29.2 & 80.7 \\
		0 & N & 0.5 & 16.8 & 79.5 & 20.3 & 80.7 \\
		0 & t & 0 & 36.4 & 64.0 & 16.4 & 21.2 \\
		0 & t & 0.5 & 13.6 & 22.5 & 10.1 & 15.6 \\ \hline
		0.5 & N & 0 & 65.8 & 99.6 & 25.9 & 70.1 \\
		0.5 & N & 0.5 & 15.4 & 66.6 & 15.9 & 67.9 \\
		0.5 & t & 0 & 39.9 & 64.9 & 20.1 & 23.4 \\
		0.5 & t & 0.5 & 18.1 & 24.7 & 12.4 & 16.8 \\ \hline \hline
		\end{tabular}
\label{table1}
\end{table}

\end{document}